\documentclass[
preprint,
showpacs,preprintnumbers,
bibnotes,
 amsmath,amssymb,
 aps,
 pra,
superscriptaddress,
longbibliography,
]{revtex4-2}

\usepackage{graphicx}
\usepackage{dcolumn}
\usepackage{bm}
\usepackage[english]{babel}
\usepackage[utf8]{inputenc}
\usepackage{geometry}
\usepackage{amsmath,amsthm,amsfonts,amssymb,amscd,mathtools}
\usepackage{hyperref}
\usepackage{lipsum}
\usepackage{enumerate}
\usepackage{textcomp}
\usepackage{fixmath}
\usepackage{xcolor}
\usepackage{float}

\begin{document}

\title{\textbf{Chiral high-harmonic generation in metasurfaces}}

\author{Piyush Jangid}
\affiliation{Nonlinear Physics Centre, Research School of Physics, Australian National University, Canberra ACT 2601 Australia}

\author{Maria Antonietta Vincenti}
\affiliation{Department of Information Engineering, University of Brescia, 25123 Brescia, Italy}

\author{Luca Carletti}
\affiliation{Department of Information Engineering, University of Brescia, 25123 Brescia, Italy}

\author{Anton Rudenko}
\affiliation{Laboratoire Hubert Curien, UMR 5516 CNRS, Universit{\'e} Jean Monnet, 42000 Saint Etienne, France}

\author{Yuri Kivshar}
\affiliation{Nonlinear Physics Centre, Research School of Physics, Australian National University, Canberra ACT 2601 Australia}

\author{Sergey Kruk}
\affiliation{ARC Centre of Excellence QUBIC, IBMD, School of Mathematical and Physical Sciences, University of Technology Sydney Ultimo, NSW 2007, Australia}
\email{Corresponding author: luca.carletti@unibs.it} \email{sergey.kruk@uts.edu.au}

\begin{abstract}
High-harmonic generation (HHG) provides the only source of attosecond pulses -- currently the shortest accessible time intervals, and it is employed as the only table-top source of light in extreme UV and soft X-ray spectral regions. Chiral HHG can be employed as an efficient tool for studying the ultrafast response of chiral properties of matter, as well as for amplifying chiroptical effects. Traditionally, chiral high harmonics were associated with gases of enantiomer molecules or, more recently, solid surfaces with helicity in their crystalline structure. Here, we bring the concept of chiral high-harmonic generation to nanophotonics, specifically to metasurfaces consisting of arrays of nanoresonators. Our system is achiral at the material as well as at the level of individual nanoresonators. Chirality rises and falls in a controlled manner via an interplay of the nanoresonator symmetry and the symmetry of the metasurface lattice. Our calculations predict high contrast in harmonic brightness between the two orthogonal circular polarizations of the pump. Our findings, at the intersection of chiral nanophotonics and strong-field optics, pave the way for chiral attosecond physics and chiral extreme UV optics in nanostructured solids.
\end{abstract}

\maketitle

\section{Introduction}
High-harmonics generation (HHG) has traditionally been associated with gases \cite{Ferray1988} and plasma \cite{Burnett1977}, and more recently it entered the realm of solid-state physics \cite{Ghimire2018}. Solid-state sources are attractive for broadening the range of HHG applications to create smaller, simpler, and cheaper systems. However, bulky solids hinder some of the key applications of HHG as the source of attosecond pulses or source of ultrashort wavelengths: short-wavelength radiation, such as extreme ultraviolet (UV) radiation, is being absorbed rapidly in the bulk solids, and ultrafast pulses are being distorted by the material dispersion.

This opens up a unique niche for HHG in ultra-thin, nanostructured solids $-$ nanoresonators \cite{Zalogina2023} and metasurfaces \cite{Jangid2024}. The ultra-thin form-factor mitigates the disadvantages of bulky solids, while the judicious design of the metasurfaces at the nanoscale can enhance the efficiency of light-matter interactions by orders of magnitude via an engineered resonant response \cite{Sergey2017,Zubyuk2021}.

Chiral metasurfaces have been employed in linear optics as well as in the generation of low-order ($2^{nd}$ and $3^{rd}$) optical harmonics.
Chirality in metasurfaces may emerge from (i) symmetry group of the material at the microscopic level, e.g., at the level of atomic lattice, (ii) symmetry group of an individual nanoresonator/unit cell, (iii) symmetry breaking by the oblique incidence of light (termed extrinsic chirality), (iv) symmetry group of a lattice of nanoresonators, (v) interplays between all above  \cite{Sergey2013,Sergey2015,Menzel2010,Berner2020,KirillReview2023,Nikitina2023,Toftul2024,Sinev2024,Toftul2025}. Both dielectric and plasmonic materials  \cite{Kauranen1998,Kauranen2005,Husu2008,Valev2010,Valev2013,Valev2018} have been explored as chiral metasurface platforms.

Nonlinear chirality of up to the third order has been demonstrated for the resonant excitation of the metasurface with circularly polarized light \cite{KirillReview2023}, to probe the maximal chiroptical response \cite{Yutao2020,Shi2022} and uncover novel applications, such as ultrafast switching of light polarizations \cite{Kang2020switching, Wang2024switching}, chiral sensing by anapole modes \cite{Serrera2024sensing}, generation of spin unlocked vortex beam \cite{Wang2024vortexbeam}, multiplexed holography \cite{Hong2022holography}, and logic gates \cite{Zhang2022logicgates}.

However, the expansion of chirality to strong-field optics and to the HHG regime, although known in gases \cite{Fleischer2014,Kfir2015} and unstructured solids \cite{Saito2017,Klemke2020,Heinrich2021}, has not yet been explored in optical nanostructures, including metasurfaces.

General links between point symmetries and harmonics generation have been established \cite{Neufeld2019,Neufeld2023}. Studies of chiral generation of multiple optical harmonics in metasurfaces, however, remain severely limited with only recent examples including theoretical calculations of direct perturbative generation of several optical harmonics. Such calculations typically retain their predictive power up to the 3rd harmonic order, with higher orders being dominated by cascaded effects and non-perturbative contributions via the generation of free carriers.


In this paper, we study numerically a nonlinear metasurface made of properly oriented achiral nanoresonators and reveal that it shows a very strong nonlinear chirality contrast. In our analysis we apply a hierarchy of methods. Firstly, a set of linear control simulations is done to identify the optimal region of geometrical parameters and resonant wavelength. This is followed by perturbative nonlinear simulations, incorporating realistic dispersion properties of bound electrons that serve as a guideline for harmonic propagation and attenuation. We consider both direct and cascaded pathways for harmonics generation. The validity of such simulations is limited due to the non-stationary nature of an ultrashort pulse (the quality of resonance might be affected by the finite number of optical cycles \cite{rudenko2018photo}), but also by pulse broadening and blueshift by laser-induced free carrier plasma \cite{sinev2021}, and the non-perturbative nature of the generated harmonics \cite{rudenko2023modeling}. We therefore finally perform non-perturbative simulations in time-domain.

Our nonlinear perturbative simulations show that nonlinear resonant chiral response is equally present for several different sources of nonlinearities, including nonlinearities arising from bulk, dipolar contributions, and nonlinearities from symmetry breaking at the surface of the resonators. Nonlinear chirality behavior is shown to emerge in the case of perturbative direct generation, perturbative cascaded generation, but also non-perturbative generation via the excitation of free-carriers for a pulse laser excitation, as short as 300 fs. Our results suggest the generation of even and odd optical harmonics down to the UV spectral range, where amorphous silicon possesses significant absorption, and they pave the way for advanced polarization control of HHG with nanofabrication technologies.

\section{Metasurface design}

\begin{figure}[hbt]
\centering
\includegraphics[width=\linewidth]{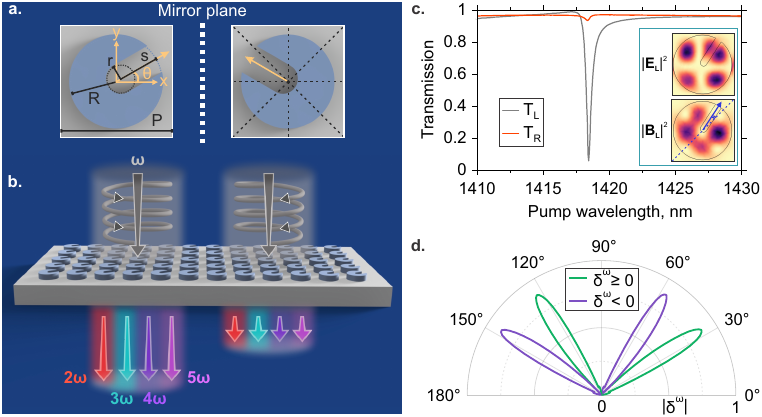}
\caption{\textbf{a.} Schematic of achiral nanoresonators arranged in a square lattice. The notations for characteristic dimensions are shown. Dashed lines are the square lattice mirror symmetry planes, and $\theta$ is the notch angle. \textbf{b.} Concept of the metasurface generating chiral high-harmonics from the second ($2\omega$) to fifth ($5\omega$) order at the incident circularly polarized pump $\omega$. \textbf{c.} Linear transmission spectrum metasurface for incident LCP and RCP, denoted as $T_L$ and $T_R$, respectively: (inset) top - electric and bottom - magnetic field density of LCP pump at the resonant wavelength. The cross-section is taken at the center of the nanoresonator. \textbf{d.} Variation of linear chirality ($\delta^{\omega}$) in the vicinity of the resonance with the notch angle.}
\label{fig:figure1}
\end{figure}

We consider an amorphous silicon (Si)-on-fused silica (SiO$_2$) metasurface consisting of achiral nanoresonators inside a square lattice, as shown in Fig. \ref{fig:figure1}b. Individual resonators are discs with notches, and the orientation of the notch with respect to the metasurface lattice is denoted by the angle of the notch $\theta$ (Fig. \ref{fig:figure1}a). When $\theta$ points along one of the lattice symmetry axes, shown as dashed lines in Fig. \ref{fig:figure1}a, the periodic array preserves the mirror symmetry and, thus, is achiral. However, for other values of $\theta$, the resonator and substrate break mirror symmetry along all geometric axes, which consequently induces chirality in metasurfaces with achiral resonators at normal incidence. In other words, the notch angle of the individual resonators acts as a knob that allows to easily control the degree of chirality of the metasurfaces.

First, we simulate numerically the linear chiral response of the metasurface with COMSOL Multiphysics by considering a lattice spacing of $P=975$~nm, a nanoresonator radius of $R=452.5$~nm, a height of $H=245$~nm, a notch size of $(r,s)=(52.5$~nm, 420~nm) and a notch angle of $\theta=57^{\circ}$ with respect to the x axis. The refractive index of silicon used in our calculations is shown in Fig. S1a in the Supplementary Information, which is consistent with the experimental data. Fig.~\ref{fig:figure1}c shows the metasurface transmission when the left circularly polarized (LCP) light and right circularly polarized light (RCP) light illuminate the metasurface at normal incidence, labeled $T_L$ and $T_R$, respectively. Fig. S2 in the Supplementary Information shows the localized electric and magnetic field densities on the resonator at resonance under LCP and RCP illumination. For metasurfaces, linear chirality is usually measured in terms of circular dichroism, which is the differential transmission of LCP and RCP \cite{Sergey2015,KirillReview2023}. However, its conventional definition involves absorption instead of transmission \cite{Atkins2005}. So we refrain from explicitly using the term circular dichroism and instead quantify linear chirality by the parameter $\delta^{\omega}$, defined as 
\begin{equation}
    \delta^{\omega} = \frac{(T_L-T_R)}{(T_L+T_R)},
\end{equation} 
which gives rise to $|\delta^{\omega}| \approx 0.88$ at the resonance. Next, we study the interplay between resonator orientation and metasurface chirality. By fine-tuning $\theta$, while keeping the other geometrical parameters the same, we observe a change in the chiral response of the metasurface ($\delta^{\omega}$): the chiral response ranges from a minimum on the lattice symmetry planes when $\theta=0^{\circ}, 45^{\circ}$ and $90^{\circ}$, to the extrema at $33^{\circ}$ and $57^{\circ}$, in the first quadrant (see Fig. \ref{fig:figure1}d). The same behavior repeats in the remaining quadrants (not shown).

\section{Nonlinear results}
\subsection{Perturbative calculations}

To analyze the nonlinear behavior of the metasurface, we first implemented a perturbative numerical model that combines the volumetric nonlinear responses and the effective second-order response that arises from symmetry breaking at the surface and bulk third-order nonlinearity. Silicon has a centrosymmetric crystal structure \cite{patnaik2003handbook}; therefore, it does not show any second-order bulk nonlinearity in either its crystalline or amorphous form. As a result, the only second-order nonlinear sources present in the resonators originate from volume and surface contributions that arise from magnetic dipoles (Lorentz force), inner-core electrons, convective nonlinear sources, and electron gas pressure \cite{adler1964nonlinear, bloembergen1968optical,scalora2019resonant, scalora2010second, hallman2023harmonic}. To take into account all these nonlinear terms, we solve the electromagnetic problem at the second-harmonic frequency as outlined in Refs. \cite{vincenti2012gain, de2013low}.

In particular, second-harmonic current density sources are calculated as the superposition of two terms: a volume term, $J_{vol}$, and a surface term, $J_{surf}$.  We then include these two currents into a finite element solver where the five harmonics are mutually coupled as described in the Supplementary Information, while the bulk third-order nonlinear susceptibility, assumed to be isotropic, is included with a dispersion profile calculated according to Miller’s rule \cite{miller1964optical, boyd2008nonlinear} (see Supplementary Information for more details on the implementation of the model). We calculate the transmitted second, third, fourth, and fifth harmonic conversion efficiencies for both LCP and RCP pumps as the transmitted power at each frequency, normalized by the incident pump power $P_{FF}$, where $P_{FF} = I_{FF}\cdot P$, where $P$ is the periodicity of the metasurface and $I_{FF} = 50$ MW/cm$^2$ is the pump irradiance. We find that contributions of higher-order nonlinearities (e.g. $\chi^{(5)}$) contribute insignificantly to the harmonics generation (see Supplementary Information Section VII and Fig. S6).

Fig. S3 in the Supplementary Information summarizes our results for linear vs. nonlinear transmission, reflection, and absorption of the metasurfaces for the two orthogonal circular polarizations, where nonlinear transmission/reflection accounts for the modulation of the epsilon by the pump (see Supplementary Information Section IV).

\begin{figure}
\centering
\includegraphics[width=0.5\linewidth]{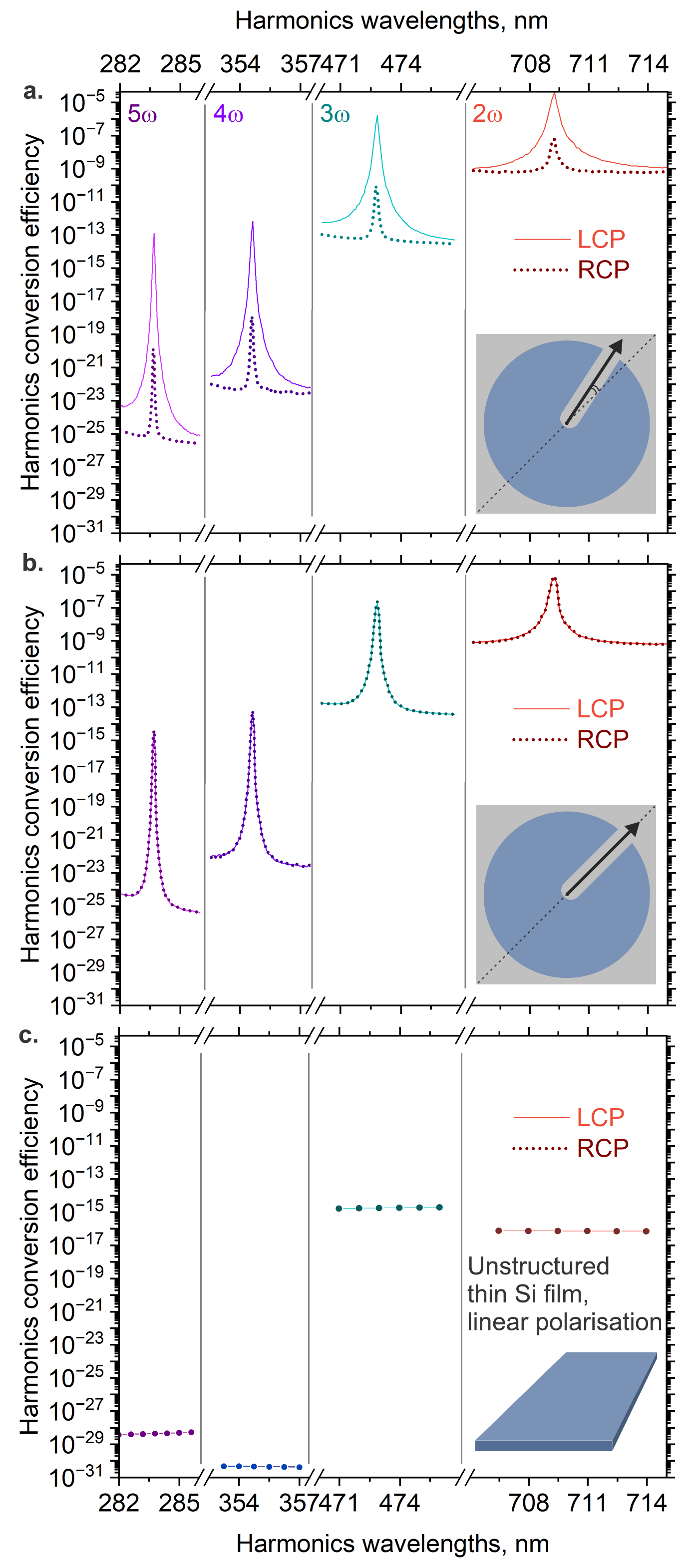}
\caption{\textbf{a.} Second-to-fifth harmonic conversion efficiencies for LCP (solid lines) and RCP (dashed lines) incident light and the notch angle of $57^{\circ}$. \textbf{b.} Same as \textbf{a.} for the notch angle of $45^{\circ}$.
\textbf{c.} Control calculation of harmonics generation from plain Si film at normal incidence for linearly polarized pump}
\label{fig:figure2}
\end{figure}

 Conversion efficiencies for both even and odd harmonics are shown in Fig. \ref{fig:figure2}a for the two cases of illumination: with LCP (solid lines) and RCP (dashed lines), and for the $\theta=57^{\circ}$ degree angle corresponding to the maximum of chirality. We contrast this calculation with Fig. \ref{fig:figure2}b where $\theta=45^{\circ}$. We stress that, although the strength of the nonlinear response for all harmonics is significantly different for the two polarizations, the second harmonic is more efficient than the third harmonic, and the fourth harmonic is stronger than the fifth harmonic, despite their inherently different origin. This can be attributed to the tensorial form of amorphous silicon that naturally suppresses odd harmonic generation under circularly polarized light illumination \cite{Tang1971} and the electric field localization profile at the resonant pump frequency that boosts second-order nonlinear processes arising from symmetry breaking at the surface of the disks (see Supplementary Information for electric and magnetic field distribution in the nanodisk resonator). 
 
 At high frequencies of optical harmonics, the metasurface unit cell size becomes larger than the harmonics' wavelengths, and the emission of optical harmonics is directed towards several diffraction orders. We calculate relative intensities of different diffraction orders for different harmonics as well as their individual polarisation states in the Supplementary Information Section VI and Figs. S4, S5.
 
 We perform a control calculation of harmonics generation from a plain Si film of thickness identical to the metasurface height, at normal incidence and for linear polarisation (see Fig. \ref{fig:figure2}c). In this case, the response is dominated by the third harmonic, which becomes brighter than the second harmonic as is conventionally expected from the bulk nonlinearities of silicon. Correspondingly, for the thin film, the fifth harmonic also becomes brighter than the fourth harmonic.


Down-conversion effects have also been included because of the high local fields that are achieved at the resonance wavelength. The impact of the feedback at the pump frequency is particularly evident for LCP excitation, with a nonlinear transmission that is modestly affected by the nonlinear interactions and nonlinear absorption that increases dramatically at the expense of nonlinear reflection. On the other hand, under RCP illumination, no significant changes are registered going from a low-intensity linear regime ($I_{FF} = 1$ W/cm$^2$) to the nonlinear regime ($I_{FF}= 50$ MW/cm$^2$).

\begin{figure}
\centering
\includegraphics[width=0.5\linewidth]{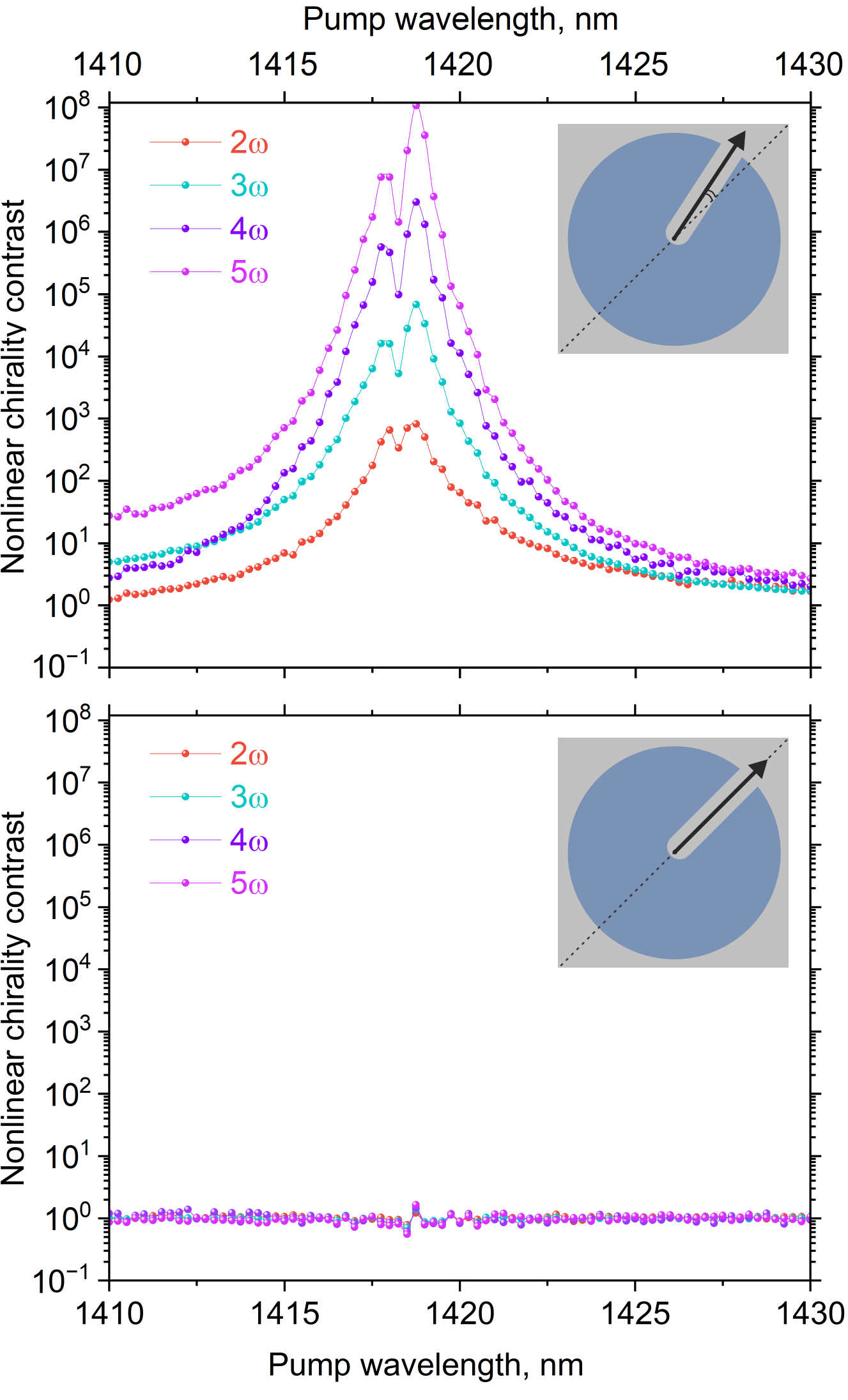}
\caption{
\textbf{a.} Contrast of harmonics generation efficiency for LCP versus RCP excitation for the notch angle of $57^{\circ}$. \textbf{b.} Same as \textbf{a.} for the notch angle of $45^{\circ}$.}
\label{fig:figure3}
\end{figure}

For a comparison of harmonics brightness between the two orthogonal circular polarizations of pump, we define nonlinear chirality contrast as

\begin{equation}
    \frac{I_L^{n\omega}}{I_R^{n\omega}}
\end{equation}

where $I_L^{n\omega}$ ($I_R^{n\omega}$) is the intensity of $n^{th}$-harmonic generation for incident LCP (RCP) pump. Our calculations predict high nonlinear chirality contrast for the optimal notch angle $\theta$ (see Fig.~\ref{fig:figure3}a). The control calculation for the $\theta$ angle corresponding to the achiral case shows no chirality contrast (see Fig.~\ref{fig:figure3}b). Interestingly, outside the resonance, we see the harmonic signal several orders of magnitude stronger than in the film. This is because the disks have a slightly higher electric field inside as compared to the film.

Finally, we calculate nonlinear chirality, quantified by $\delta^{n\omega}$, as

\begin{equation}
    \delta^{n\omega}=\frac{I_L^{n\omega}-I_R^{n\omega}}{I_L^{n\omega}+I_R^{n\omega}}
\end{equation}

The $\delta^{n\omega}$ at resonance is $>99.9\%$ for 4$^{th}$ and 5$^{th}$ harmonics, $>99.7\%$ for 2$^{nd}$ and 3$^{rd}$ harmonics. To provide evidence that the chiral nonlinear dynamics is enabled by the specific notch angle, we compare the chirality contrast of the studied metasurface with the one from a metasurface with $\theta=45^{\circ}$ and get the achiral response in the latter case.

\subsection{Non-perturbative calculations}

We next explore the non-perturbative regime of optical harmonic generation. When materials are exposed to intense ultrashort laser pulses, their nonlinear optical response is altered due to the photo-excitation of electron-hole pairs. In addition, the resonant behavior of the metastructure is affected by swift changes in transient optical properties of the electron plasma tightly confined within a nanoscale hotspot inside a subwavelength resonator. Under such conditions, the spectral output of both low- and high-order harmonics becomes highly sensitive to the presence of laser-induced carriers.

\begin{figure}
\centering
\includegraphics[width=0.99\linewidth]{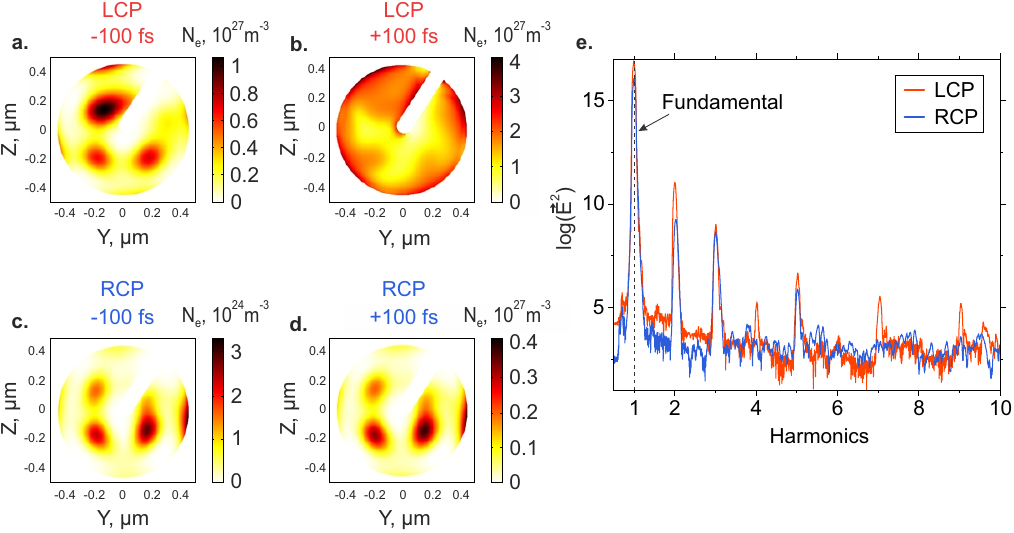}
\caption{Spatial distribution of free electrons in one nanoresonator under the LCP (\textbf{a.}, \textbf{b.}) versus RCP (\textbf{c.}, \textbf{d.}) for the optimal geometry. The density is integrated over the disk height and shown at the time $-100$ fs before and $100$ fs after the laser pulse peak. \textbf{e.} High harmonics spectra are compared for the LCP and RCP.}
\label{fig:figure4}
\end{figure}

In what follows, we solve 3D full-vector nonlinear Maxwell equations where the ultrashort laser pulse propagation is modeled within finite-difference time-domain (FDTD) approach coupled with a rate equation for conduction band electrons in laser-excited silicon. Simulations include an intraband nonlinear current, responsible for odd-order harmonics, and nonlocal sources, such as Lorentz force, a convection term, and a quadrupolar term, responsible for even harmonics. The details are given in the Supplementary Information Section VIII. Numerical results for a given geometry with the optimal parameters (including the resonant wavelength $1418$ nm in linear simulations and the notch angle of $57^{\circ}$) and a laser pulse duration of $300$ fs are shown in Fig. \ref{fig:figure4}. We consider an infinite array of periodic meta-atoms by setting periodic boundary conditions in {Y} and {Z} directions.  

The carrier dynamics illustrated in Fig. \ref{fig:figure4} for the intensity $I \approx 1$ GW/cm$^2$ indicates the following non-intuitive behavior. At earlier stages of laser excitation (a,c), free electrons are mainly localized within three hot spots, non-identical for LCP and RCP, with a stronger electric field enhancement for LCP, and consequently, a higher number of generated carriers. This indicates that the resonant behavior starts to manifest even after a first dozen of optical cycles. If the intensity is not strong enough, as in the case of RCP, the pattern remains unchanged even after the pulse peak in Fig. \ref{fig:figure4}d, and the free carriers do not affect significantly the pulse propagation and spatial energy deposition. In this case, the harmonic spectra of the transmitted laser pulse in Fig. \ref{fig:figure4}e (blue line) clearly show only the second, third, and fifth harmonics above the noise floor. However, if the field enhancement is stronger and the density of free carriers is higher, as in this case for LCP, the transient optical properties start to play a role by shaping the spatial distribution of plasma inside meta-atom. At a certain stage, there is transition towards almost homogeneous electron plasma distribution inside meta-atom as in Fig. \ref{fig:figure4}(b). The transmission harmonic spectra, in addition to low order harmonics observed for RCP, show features of high-order harmonics, including seventh and ninth, but also the fourth harmonic. On the other side, the ratio between LCP and RCP of lower order harmonics is much less than predicted by perturbative simulations, suggesting that the efficiency of nonlinear emission might be affected if the meta-atom switches off-resonance by strong laser excitation. This also indicates that there might be an optimal intensity and pulse duration to obtain the strongest nonlinear chirality in laser-excited nanostructures.  

\section{Conclusion}

Our study provides a theoretical demonstration of the chiral generation of multiple optical harmonics, including the harmonics of both even and odd orders within perturbative and non-perturbative approaches. Both volumetric and surface nonlinearities were factored. We further studied direct vs. cascaded generation pathways. Chiral harmonics in our system arise from the intricate interplay between the point symmetries of the unit cell and those of the overall lattice structure of the metasurface. By carefully engineering these symmetries at the nanoscale, we have established a robust mechanism for inducing chirality in nonlinear optical processes, even in materials that are intrinsically achiral at the atomic level. Our findings reveal the fundamental role of nanoscale structures in tailoring and enhancing chiral responses in strong-field light-matter interactions, offering a powerful alternative to relying solely on naturally chiral materials, which may be limited in availability or optical performance. Instead, our results suggest that conventional material platforms can be transformed into highly efficient chiral optical media through precise nanostructuring. Beyond its fundamental significance, our work paves the way towards practical applications of chiral photonics, including ultrafast spectroscopy and the development of novel light sources with tailored polarization properties. Moreover, the ability to achieve strong chiral effects through structured solids opens exciting opportunities in many areas such as quantum optics, optical information processing, and the control of light-matter interactions at extreme field intensities.

\section*{Acknowledgements}
The authors thank Ivan Toftul for his highlighting comments. SK acknowledges financial support from the Australian Research Council (grant DE210100679). MAV and LC acknowledge financial support from the NATO Science for Peace and Security program (Grant no. 5984). YK was supported by the Australian Research Council (Grant No. DP210101292) and the International Technology Center Indo-Pacific (ITC IPAC) via Army Research Office (contract FA520923C0023).

\section*{Supplementary Information}
See Supplementary Information for supporting content.

\clearpage    
\thispagestyle{empty}    
\renewcommand{\thefigure}{S\arabic{figure}}
\setcounter{figure}{0}

\begin{center}
	{\Large \bfseries Supplementary Information: Chiral high-harmonics generation in metasurfaces}     
\end{center}

\newpage
\section{Linear Permittivity Response}
\begin{figure}[htb!]
\centering
\includegraphics[width=\linewidth]{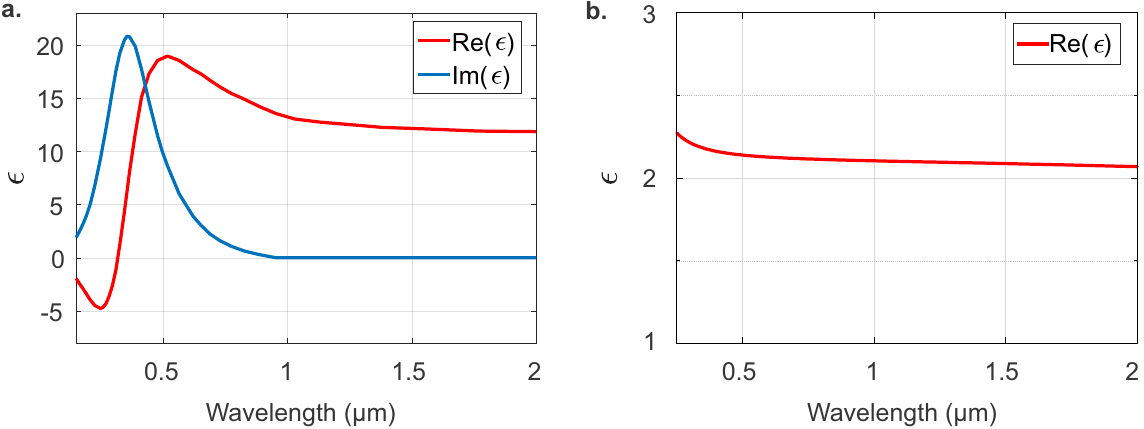}
\caption{Permittivity of \textbf{a.} amorphous silicon, and \textbf{b.} fused silica}
\label{fig:figS1}
\end{figure}

The experimental data of amorphous silicon (Ref.~\cite{Pierce1972}) and fused silica (Ref.~\cite{Malitson1965}) are shown in Fig. \ref{fig:figS1}. The extinction coefficient of fused silica is considered zero ($Im(\epsilon)=0$). 

\section{Electric and Magnetic Field Distribution}
\begin{figure}[!htbp]
\centering
\includegraphics[width=0.75\linewidth]{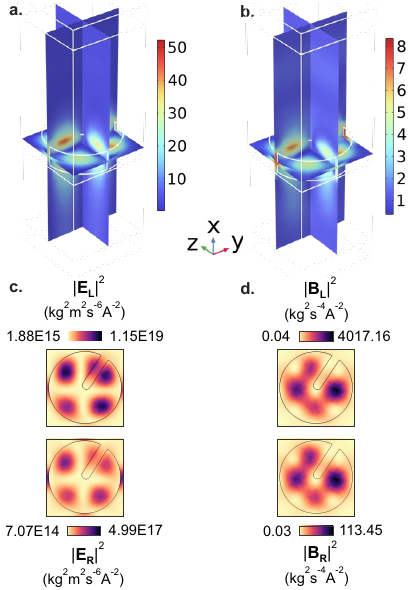}
\caption{Electric field density of LCP (\textbf{a.}, \textbf{c.} (top)) and RCP (\textbf{b.}, \textbf{c.} (bottom)) pump at the resonant wavelength 1418 nm. \textbf{d.} Magnetic field density for LCP (top) and RCP (bottom) pump at the resonant wavelength. The cross-section is taken at the center of the nanoresonator.}
\label{fig:figS2}
\end{figure}

Electric field enhancement, calculated as ${|{E}|}/{E_0}$, where $E_0$ is the incident electric field calculated at the resonance wavelength $\lambda = 1418$ $nm$ for LCP (Fig. \ref{fig:figS2} a) and RCP (Fig. \ref{fig:figS2} b). We note that the electric (Fig. \ref{fig:figS2} a,b,c) and magnetic (Fig. \ref{fig:figS2} d) field densities are strongly localized at the surface of the nanodisks, boosting even-harmonic generation that arises from symmetry breaking at the interfaces.

\section{Linear vs Nonlinear Transmission}

\begin{figure}[H]
\centering
\includegraphics[width=\linewidth]{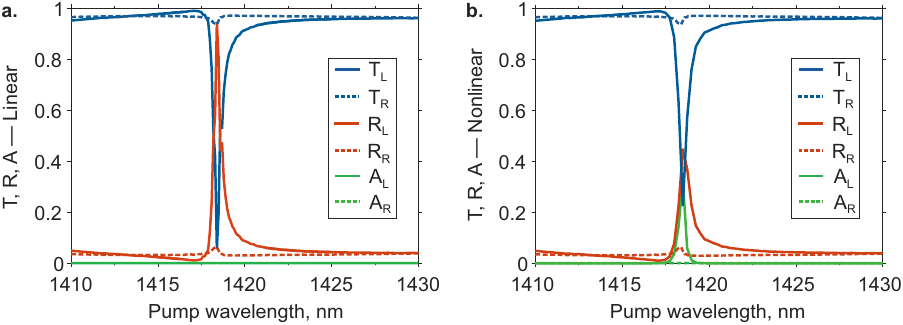}
\caption{\textbf{a.} Linear (pump intensity $I_{FF}$ = 1 W/cm$^2$) and \textbf{b.} nonlinear (pump intensity $I_{FF}$ = 50 MW/cm$^2$) transmission (T), reflection (R), and absorption (A) for LCP (L) and RCP (R) incident light.}
\label{fig:figS3}
\end{figure}

\section{Second-order nonlinear response}

Second-order nonlinear currents are linked to the fundamental frequency (FF) electric field and to the material parameters as follows:

\begin{align}
    \mathbf{\hat{n}\cdot J_{s}}=&i\frac{n_0e^3}{2m^{*}_b}\frac{3+\varepsilon_{FF}}{(\omega+i\gamma_0)^2(2\omega+i\gamma_0)}E^2_{FF,\perp}\\
    \mathbf{\hat{t}\cdot J_{s}}=&i\frac{n_0e^3}{2m^{*}_b}\frac{1}{(\omega+i\gamma_0)^2(2\omega+i\gamma_0)}E_{FF,\perp}E_{FF,\parallel}\\
    \mathbf{J_{v}}=&i\frac{n_0e^3}{2m^{*}_b}\frac{1}{\omega(\omega+i\gamma_0)(2\omega+i\gamma_0)}\nonumber\\
    &\left[\frac{\gamma_0}{\omega+\gamma_0}(\mathbf{E_{FF}\cdot\nabla})\mathbf{E_{FF}}-\frac{i}{2}\mathbf{\nabla}(\mathbf{E_{FF}\cdot E_{FF}})\right]
\end{align}

where $n_0=\varepsilon_0 m_b^{*2}\omega_p^2/e^{2}$ is the bound electrons density, the effective bound electron mass is assumed equal to the electron mass $m_b^{*}=m_e=9.10938188\times10^{-31}$~kg, $e$ is the elementary charge, and the plasma frequency, $\omega_p$, and decay rate, $\gamma_0$, used to fit the dielectric permittivity of amorphous silicon are $\omega_p=1.6953\times10^{16}$~rad/s and $\gamma_0=2.4487\times10^{15}$~rad/s, respectively. The value of the permittivity $\varepsilon_{FF}$ of Si at the FF, $\omega$ is the angular frequency of the FF field, $\mathbf{E_{FF}}$ is the FF electric field phasor, and $\mathbf{\hat{n}}$ and $\mathbf{\hat{t}}$ are unit vectors pointing in directions outward normal and tangential to the nanodisk surface, respectively. Moreover, $\mathbf{E_{FF,\perp}}$ and $\mathbf{E_{FF,\parallel}}$ are the normal and tangential components of the FF electric field in the local boundary coordinate system defined by $\mathbf{\hat{n}}$ and $\mathbf{\hat{t}}$, and are evaluated inside the Si resonator regions.

\section{Implementation of the five harmonics model}
We developed a five-harmonics model to evaluate the conversion efficiencies of the metasurface made of Si nanodisks. To simplify the computational approach we calculate two different dispersion profiles for the third-order nonlinear susceptibility tensor, one that catches the nonlinear processes occurring at frequencies between the pump and the second harmonic described by $\chi^{(3)_{i,j,k,l}}(\omega;\omega,\omega,-\omega)$, while the other is associated with processes occurring from the third to the fifth harmonics described by $\chi^{(3)_{i,j,k,l}}(3\omega;\omega,\omega,\omega)$ where $ijkl$ are cartesian axes. Since we assume the nanodisk to be made of amorphous silicon, we consider tensorial elements corresponding to an isotropic response (Ref.~\cite{boyd2008nonlinear}). Because of the high field localization values reached in the silicon resonators, we also include pump depletion, self- and cross-phase modulation processes between all harmonics, discarding those terms that contain powers of the second or higher harmonics. This results into the following nonlinear polarization terms at the pump and harmonic frequencies:

\begin{align}
P^{NL}_{\omega,i}=&\chi^{(3)}_{ijkl}\left[(3E_{\omega,j}|E_{\omega,j}|^2+2E_{\omega,j}|E_{\omega,k}|^2+2E_{\omega,j}|E_{\omega,l}|^2+E^*_{\omega,j}\left(E^2_{\omega,k}+E^2_{\omega,l}\right)\right]\\
P^{NL}_{2\omega,i}=&\chi^{(3)}_{ijkl}\left[6E_{2\omega,j}|E_{\omega,j}|^2+2E_{2\omega,j}|E_{\omega,k}|^2+2E_{2\omega,j}|E_{\omega,l}|^2\right]+\nonumber\\
&\chi^{(3)}_{ijkl}\left[2E_{2\omega,k}\left(E^*_{\omega,j}E_{\omega,k}+E^*_{\omega,k}E_{\omega,j}\right)+2E_{2\omega,l}\left(E^*_{\omega,j}E_{\omega,l}+E^*_{\omega,l}E_{\omega,j}\right)\right]+\nonumber\\
&\chi^{(3)}_{ijkl}\left[E_{4\omega,j}\left(3E^{*2}_{\omega,j}+E^{*2}_{\omega,k}+E^{*2}_{\omega,l}\right)+2E^{*}_{\omega,j}\left(E_{4\omega,k}E^{*}_{\omega,k}+E_{4\omega,l}E^{*}_{\omega,l}\right)\right]\\
P^{NL}_{3\omega,i}=&\chi^{(3)}_{ijkl}E_{\omega,j}\left(E_{\omega,j}^2+E_{\omega,k}^2+E_{\omega,l}^2\right)\\
P^{NL}_{4\omega,i}=&\chi^{(3)}_{ijkl}\left[E_{2\omega,j}\left(3E^2_{\omega,j}+E^2_{\omega,k}+E^2_{\omega,l}\right)+2E_{\omega,j}\left(E_{2\omega,k}E_{\omega,k}+E_{2\omega,l}E_{\omega,l}\right)\right]+\nonumber\\
&\chi^{(3)}_{ijkl}[E_{4\omega,k}\left(6|E_{\omega,j}|^2+2|E_{\omega,k}|^2+2|E_{\omega,l}|^2\right)+\nonumber\\
&\quad\quad\,\, 2E_{4\omega,k}\left(E^*_{\omega,j}E_{\omega,k}+E^*_{\omega,k}E_{\omega,j}+E^*_{\omega,j}E_{\omega,l}+E^*_{\omega,l}E_{\omega,j}\right)]\\
P^{NL}_{5\omega,i}=&\chi^{(3)}_{ijkl}\left[E_{3\omega,j}\left(3E^2_{\omega,j}+E^2_{\omega,k}+E^2_{\omega,l}\right)+E_{\omega,j}(2E_{3\omega,k}E_{\omega,k}+2E_{3\omega,l}E_{\omega,l})\right]
\end{align}

\section{Diffraction of generated harmonics}
\begin{figure}[!htbp]
\centering
\includegraphics[width=\linewidth]{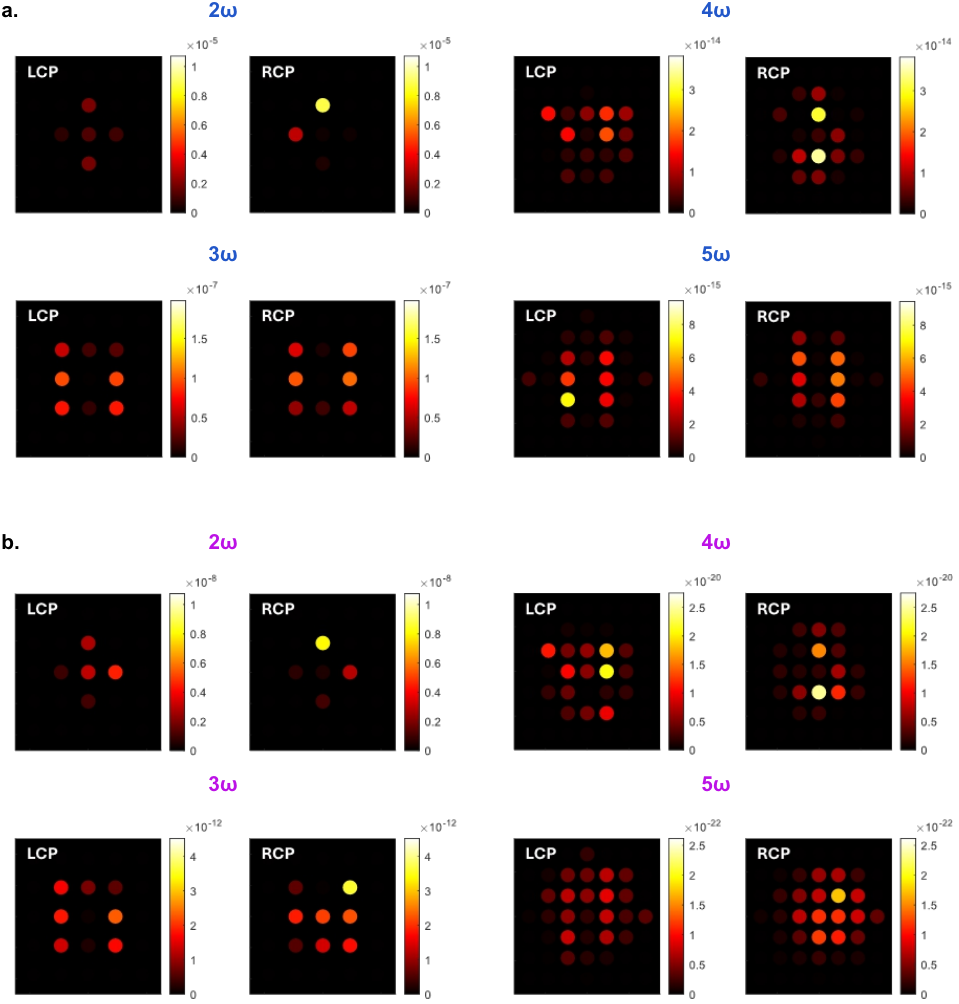}
\caption{Far-field emission in NA = 0.9 for \textbf{a.} LCP and \textbf{b.} RCP pump.}
\label{fig:figS4}
\end{figure}

The wavelength of the generated harmonics is shorter than the period of the metasurface, resulting in light emission into multiple diffraction channels. The calculations presented in this work account for the total emission by summing contributions from all diffraction channels at each harmonic order. This is performed by calculating the flux of the outward Poynting vector with respect to a plane parallel to the metasurface plane inside the substrate. Moreover, we fully characterize the emission of the harmonics in the different diffraction channels for both left- and right-handed circular polarization excitation. The results, shown in Fig.~\ref{fig:figS4}, reveal how the harmonics' emission is always distributed in several diffraction orders, with the efficiencies associated with RCP excitation always being orders of magnitude lower than the LCP pump. We note that circular dichroism is preserved across all harmonics and for all diffraction orders.

\begin{figure}[htb]
\centering
\includegraphics[width=\linewidth]{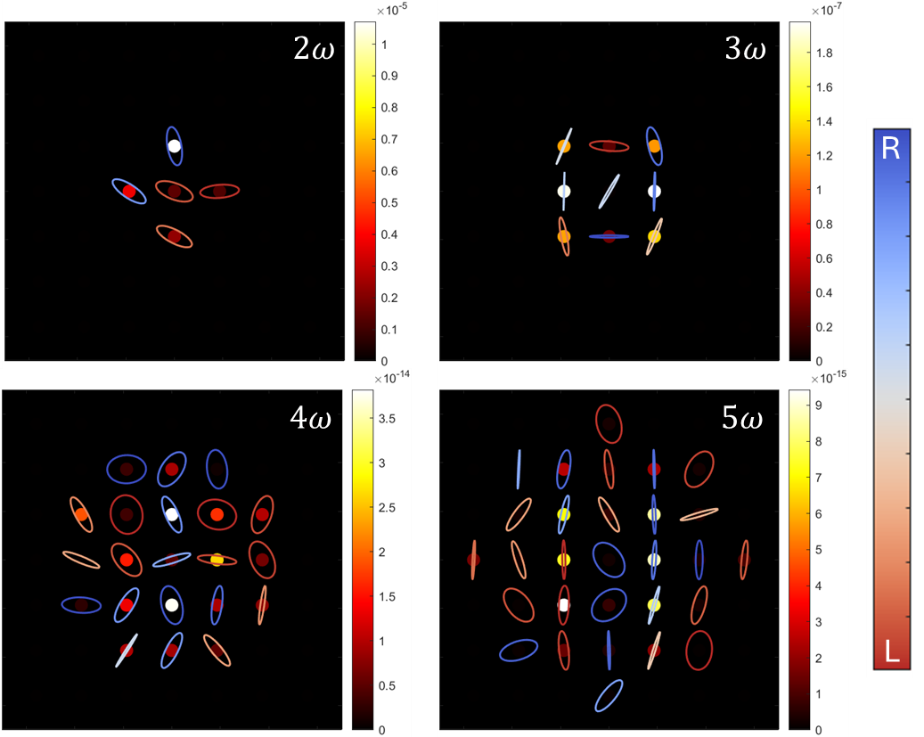}
\caption{Far-field radiation of harmonics with the relative polarization state for all diffraction orders. Far-field emission is calculated for NA = 0.9 and LCP pump.}
\label{fig:figS5}
\end{figure}

Further, we observe that the polarization state of the generated harmonics varies in different diffraction channels and is generally elliptical. For example, if we consider the $(0,0)$ diffraction order, we observe that the chirality of second-harmonic emission is left, while it is right for all other harmonics. Fig.~\ref{fig:figS5} shows the far-field radiation of different harmonics generated by the metasurface when the pump is LCP and tuned to the quasi-BIC wavelength. The filled circles represent the diffraction order with color-coded efficiency calculated as the ratio between the power in the diffraction order and the pump power (the same as in Fig.~\ref{fig:figS4}a). Ellipses depict the polarization state with color-coded chirality.

\section{Direct vs cascaded generation of optical harmonics}

We compare here the difference between the model based on $\chi^3$  and an extended version of the model where $\chi^5$  contributions are included for 3rd and 5th harmonic generation and feedback process at the pump frequency. More specifically, we are extending our equation sets for pump, third, and fifth harmonic generation 

from:

\begin{align*}
\mathbf{P}^{NL}(\omega) &= \varepsilon_0 \left[ \chi^{(3)} |\mathbf{E}_\omega|^2 \mathbf{E}_\omega \right] \\
\mathbf{P}^{NL}(3\omega) &= \varepsilon_0 \left[ \chi^{(3)} \mathbf{E}_\omega \mathbf{E}_\omega \mathbf{E}_\omega \right] \\
\mathbf{P}^{NL}(5\omega) &= \varepsilon_0 \left[ \chi^{(3)} \mathbf{E}_{3\omega} \mathbf{E}_\omega \mathbf{E}_\omega \right]
\end{align*}

to:

\begin{align*}
\mathbf{P}^{NL}(\omega) &= \varepsilon_0 \left[ \chi^{(3)} |\mathbf{E}_\omega|^2 \mathbf{E}_\omega + \chi^{(5)} |\mathbf{E}_\omega|^4 \mathbf{E}_\omega \right] \\
\mathbf{P}^{NL}(3\omega) &= \varepsilon_0 \left[ \chi^{(3)} \mathbf{E}_\omega \mathbf{E}_\omega \mathbf{E}_\omega + \chi^{(5)} |\mathbf{E}_\omega|^2 \mathbf{E}_\omega \mathbf{E}_\omega \mathbf{E}_\omega \right] \\
\mathbf{P}^{NL}(5\omega) &= \varepsilon_0 \left[ \chi^{(3)} \mathbf{E}_{3\omega} \mathbf{E}_\omega \mathbf{E}_\omega + \chi^{(5)} \mathbf{E}_\omega \mathbf{E}_\omega \mathbf{E}_\omega \mathbf{E}_\omega \mathbf{E}_\omega \right]
\end{align*}

Since we do not have any bulk second and fourth order susceptibilities, second and fourth harmonics will be affected by the changes introduced at the pump frequency through cascading terms that are already present in the original model. $\chi^5$  values are estimated by means of an atomic field scaling approach, which provides a useful dimensional estimate based on the characteristic atomic field strength.

\begin{figure}[!htbp]
\centering
\includegraphics[width=\linewidth]{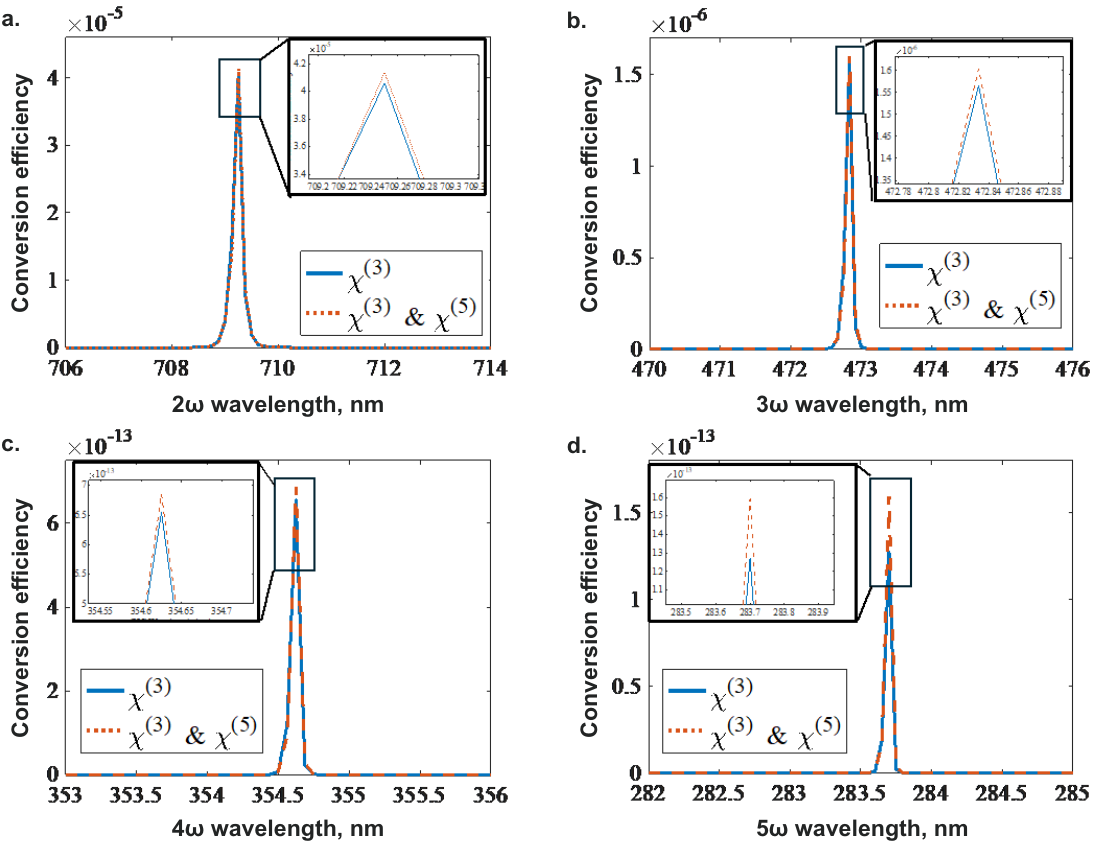}
\caption{\textbf{a.} Second, \textbf{b.} third, \textbf{c.} fourth, and \textbf{d.} fifth harmonic conversion efficiency calculated by including only $\chi^3$ (blue, solid lines in all plots) and then including both $\chi^3$ and $\chi^5$) (red, dashed lines in all plots). To appreciate the slight increase on all conversion efficiencies due to $\chi^5$ inclusion zoom ins are provided as insets in all plots.}
\label{fig:figS6}
\end{figure}

Our results, shown in Fig. \ref{fig:figS6}, demonstrate that $\chi^5$ has a negligible impact on the harmonic yield for the structure under study. It goes without saying that including $\chi^7$ contributions will have an even lower impact on the harmonic processes up to the 5th order.

\section{Non-perturbative model}

Full-vector Maxwell equations with a current density for free carriers $\vec{J_e}$ and $\vec{J_{Kerr}}$ for Kerr effect are solved as follows:    
\begin{align} \begin{cases} $$ \label{Maxwell}	\displaystyle{\frac{\partial{\vec{E}}}{\partial{t}}=\frac{\nabla\times\vec{H}}{\epsilon_0}-\frac{(\vec{J_e}+\vec{J_{Kerr}})}{\epsilon_0}}\\		
	\displaystyle{\frac{\partial{\vec{H}}}{\partial{t}}=-\frac{\nabla\times\vec{E}}{\mu_0}},
$$ \end{cases} \end{align}
where $\vec{E}$ and $\vec{H}$ are the electric and the magnetic fields, $\epsilon_0$ and $\mu_0$ are the permittivity and the permeability of free space. Kerr nonlinearity is $n_2 = 4.5\cdot{10}^{-14} {cm}^2/W$ at $\lambda = 1.4$ ${\mu}m$ (Ref.~\cite{bristow2007}) and the corresponding current reads as $\vec{J}_{Kerr} = \frac{4n_{2}\epsilon_0^{3/2}}{3\sqrt{\mu_0}}\frac{\partial}{\partial{t}}\left[\vec{E}(\vec{E}\cdot\vec{E})\right]$. The conduction band current is related to the polarization $\vec{P}$ as $\vec{J_e} = \frac{\partial{\vec{P}}}{\partial{t}}$. Constant refractive indices were used for fused silica $n_{SiO_{2}} = 1.4214$ at $1.42$ ${\mu}m$ and $n = 1$ for vacuum. Due to a much larger bandgap than silicon, it is safe to assume that the free carrier density is negligible inside the fused silica substrate.  

The electric field source is a circularly polarized Gaussian pulse with a full width at half maximum of $\vartheta = 300$ fs and a central wavelength of $\lambda = 1418$ $nm$. Light propagates along the $x$ direction. Spatially, we consider a plane wave. Periodic boundary conditions are set in the $y$ and $z$ directions to introduce a periodic array of identical meta-atoms.

Maxwell equations are solved by the finite-difference time-domain (FDTD) approach with convolutional perfect matched layers (CPML) at the boundaries \cite{taflove1995}. The electron current density $\vec{J_e}$ is coupled to Maxwell equations (\ref{Maxwell}) by applying the auxiliary differential equation technique.  

The equations describing the optical response of carriers in the conduction band, the involved photo-ionization processes, and the nonlocal effects in the electronic fluid (Lorentz force, quadrupolar and convection terms) are written as follows 
\begin{align} \begin{cases} \label{Conduction}$$ 
\displaystyle{\frac{\partial{\vec{J}_e}}{\partial{t}} = - {\vec{J}_e}{\nu_e} +\frac{e}{m_e^*}\left[e{N_e}\vec{E}+\vec{J}\times{\mu_0}\vec{H} + (\vec{P}\cdot{\nabla})\vec{E}\right] - \sum_k \frac{\partial{\left(\vec{J}J_k/e{N_e}\right)}}{\partial{r_k}}} \\
\displaystyle{\frac{\partial{N_e}}{\partial t} =
    \frac{N_a-N_e}{N_a}w_{PI}(|\vec{E}|) -
    \frac{C_A{N_e}^3}{C_A\tau_{rec}N_e^2+1}},\\
$$ \end{cases} \end{align}
where $e$ and $N_e$ are the electron charge and density, $m_e^* = 0.18m_e$ is the reduced electron-hole mass and $\nu_e = 10^{15}$ s$^{-1}$ is the electron collision frequency related to the Drude damping time. $w_{PI}$ is the Keldysh photo-ionization rate calculated for the corresponding wavelength $\lambda$ and with direct Si bandgap of $3.4$ eV, $N_a = 5\times{10}^{22}$~cm$^{-3}$ is the Si atom density, $C_A = 3.8\times{10}^{-31}$~cm$^6$/s is the Auger recombination rate, $\tau_{rec} = 6\times{10}^{-12}$~s is the minimum Auger recombination time (Ref.~\cite{yoffa1980}).

\bibliography{bibliography}

\end{document}